\documentclass[10pt, conference, compsocconf]{IEEEtran}
\usepackage{graphicx}
\usepackage{multirow}
\usepackage{multicol}
\usepackage{makecell}
\usepackage{tabularx}
\usepackage{url}
\ifCLASSINFOpdf
\else
\fi
\begin{document}
	\title{Evaluation of Trace Alignment Quality and its Application in Medical Process Mining}
	
	\author{\IEEEauthorblockN{Moliang Zhou, Sen Yang, Xinyu Li, Shuyu Lv, Shuhong Chen, Ivan Marsic}
		\IEEEauthorblockA{Department of Electrical and Computer Engineering\\
			Rutgers University\\
			Piscataway, New Jersey, USA\\
			Email: \{mz330, sy358, xl264, sl1316, sc1624, marsic\}@rutgers.edu}
		\and
		\IEEEauthorblockN{Richard Farneth, Randall Burd}
		\IEEEauthorblockA{Division of Trauma and Burn Surgery\\Children's National Medical Center\\
			Washington, D.C., USA\\
			\{rfarneth, rburd\}@childrensnational.org}
	}
	\maketitle
	
	\begin{abstract}
		Trace alignment algorithms have been used in process mining for discovering the consensus treatment procedures and process deviations. Different alignment algorithms, however, may produce very different results. No widely-adopted method exists for evaluating the results of trace alignment. Existing reference-free evaluation methods cannot adequately and comprehensively assess the alignment quality. We analyzed and compared the existing evaluation methods, identifying their limitations, and introduced improvements in two reference-free evaluation methods. Our approach assesses the alignment result globally instead of locally, and therefore helps the algorithm to optimize overall alignment quality. We also introduced a novel metric to measure the alignment complexity, which can be used as a constraint on alignment algorithm optimization. We tested our evaluation methods on a trauma resuscitation dataset and provided the medical explanation of the activities and patterns identified as deviations using our proposed evaluation methods.
	\end{abstract}
	

	\section{Introduction}\label{sec_1}
	\subsection{Motivation}
\begin{figure}[!b]
		\begin{center}
			\includegraphics[width=2.5in]{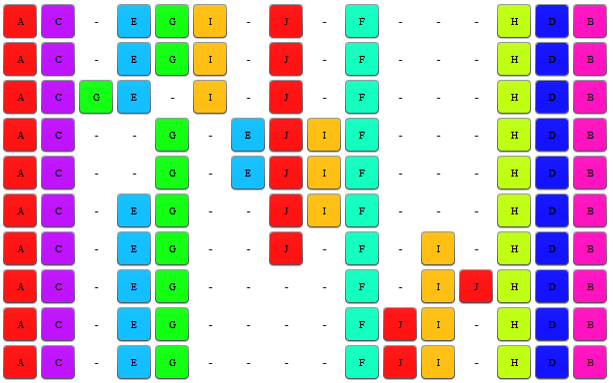}
		\end{center}
		\caption{An example of trace alignment. Activities of the same type are aligned in the same column, and each row represents an individual case formed by a trace of activities in chronological order}
		\label{fig_intro}
	\end{figure}

	Process mining has proven useful in the various domains, including clinical and administrative processes in health care \cite{rojas2016process}. A medical process mining application is the modeling of chest pain treatments and healthcare delivery \cite{partington2015process}. Trace alignment is an algorithm used in process mining to discover common work patterns and deviations from the established practice. Trace alignment takes as input sequences of activities performed during different process executions (i.e., ``trace''), and finds for each activity finds the best match in different traces \cite{rozinat2007towards}. The alignment result is a matrix where the rows represent traces, and the same type activities are aligned in the same columns (\figurename{\ref{fig_intro}}). The activities most commonly executed in a similar chronological order form a ``consensus sequenc'' \cite{bose2010trace}\cite{bouarfa2012workflow}\cite{yang2016duration}. The trace alignment algorithm originates from bioinformatics, where it is used to align protein and gene sequences to identify common structures and mutations. The parameters for alignment algorithms, however, are determined subjectively or based on expert assumption instead of automatic adjustment \cite{wu2000some}. Proper evaluation of the alignment result is therefore essential for generating an accurate alignment \cite{rozinat2007need}, and the evaluation results can help optimize the alignment algorithm parameters.
	
	The alignment accuracy measures how many activities and patterns in the alignment are aligned correctly, which determines the alignment’s ability to extract useful knowledge and insights about the process \cite{rozinat2007towards}. The correct alignment of activities and patterns means whether the aligned activities or pairs (1) are the same as in the reference or (2) satisfy certain preset constraints. The alignment accuracy could be measured from different aspects: whether the activities and patterns are accurately aligned, whether the consensus activities and potential deviations can be identified in the alignment result, and whether dissimilar activities are misaligned in the same column.
	
	Currently, methods based on the sum-of-pairs score are the most commonly used for accuracy evaluation \cite{gonnet2000evaluation}\cite{setubal1997introduction}. While the reference-based sum-of-pairs score approach \cite{gonnet2000evaluation} requires having a reference alignment to calculate the number of aligned pairs in the alignment result, the reference-free sum-of-pairs score approach \cite{setubal1997introduction} does not need the reference alignment but assumes that only the activities of the same type can be aligned. Accuracy evaluation methods do not perform well for more complex patterns in the alignment instead of activity pairs \cite{rozinat2007need}, particularly in scenarios where the reference alignment (ground truth) is unavailable.
	
	\begin{table*}[!t]
		\renewcommand{\arraystretch}{1.3}
		\caption{Summary of the current evaluation methods}
		\label{table_currentsummary}
		\centering
		\begin{tabular}{|c|c|c|c|}
			\hline
			Current Evaluation Methods & Reference Required &Attributes&Limitation\\
			\hline
			\hline
			Reference-based sum-of-pairs score \cite{gonnet2000evaluation}& Yes&Activities type&/\\
			\hline
			Reference-free sum-of-pairs score \cite{setubal1997introduction}& No&Activities type&/\\
			\hline
			Column score \cite{thompson1999comprehensive}& Yes&Activities type&Extremely sensitive to alignment errors\\
			\hline
			Misalignment score \cite{bose2012process} & No&Activities type, patterns of activities&High dependency on the pattern chosen\\
			\hline
			Information score \cite{bose2012process} & No&Activities type, activities frequency&Limited to a single column\\
			\hline
		\end{tabular}
	\end{table*}

	Misalignment score is a reference-free method for evaluating alignment accuracy \cite{bose2012process}. It measures alignment quality by checking if certain patterns are aligned. Other evaluation methods for trace alignment are adopted from molecular sequence alignment, including column score and sum-of-pairs score \cite{thompson1999comprehensive}. However, biological sequences unlike process traces, are subject to known restrictions or contain known structures, and have reference alignments established by domain experts for evaluation \cite{edgar2004comparison}\cite{de2012multi}\cite{thompson1999balibase}. In process mining, especially in the early stage of discovering common activity patterns and deviations, activity restrictions or structures are unknown, and establishing a reference alignment is laborious and requires domain expertise. In addition, as new traces are collected over time, continuous updating of the reference alignment is not practical. Because current trace alignment approaches focus on acquiring knowledge without a reference alignment, reference-free alignment evaluation methods are preferred \cite{de2012multi}. 
	
	Another metric of alignment quality is alignment confidence, which distinguishes the consensus activity from other activities in the column. The confidence can be measured by the information score which quantifies the information entropy of a single column based on the activity type and the frequency of non-empty elements of the column \cite{bose2012process}.
	 
	Current evaluation methods suffer from various limitations (\tablename{\ref{table_currentsummary}}): the column score is sensitive to the alignment error; the misalignment score is determined by arbitrarily chosen patterns and does not reflect the overall alignment quality; and the information score only evaluates individual columns. To address these limitations, we modified the misalignment score and information score. In addition, we introduced a novel metric of alignment complexity, which measures the redundancy in the alignment based on the alignment's length. We then integrated the existing and our new metrics for evaluating trace alignment including alignment accuracy, confidence, and complexity, which can reflect the overall quality of the alignment result without a reference alignment. 
	
	We validated our evaluation methods on the alignments of medical process data. Our method outperformed previous evaluation methods in terms of the correlation between the evaluation results and the number of alignment errors. In addition, we provided medical explanations of the process deviations identified by our new evaluation methods.

	\subsection{Related Work}
	Two widely used alignment accuracy evaluation methods are sum-of-pairs score (SPS) and column score (CS) \cite{thompson1999balibase}\cite{setubal1997introduction}\cite{gonnet2000evaluation}. There are two versions of the sum-of-pairs score: reference-based \cite{setubal1997introduction} and reference-free \cite{gonnet2000evaluation}. Reference-based sum-of-pairs score compares the aligned activities with a reference alignment to calculate the number for correctly aligned activity pairs. Reference-free sum-of-pairs score considers two activities as correctly aligned if they are of the same type (match). Column score compares the columns in the alignment result to the reference alignment to check if the columns are correctly aligned.
	
	Another method for evaluating alignment accuracy is misalignment score, which measures the distance between the incorrectly aligned instances of a pattern within the traces. The distance is defined as the number of columns between two activities that are supposed to be aligned in the same column. The details of the misalignment score will be discussed later. The misalignment score depends on the pattern chosen and is considered as a pattern-wise alignment accuracy \cite{bose2012process}.
	
	Alignment confidence is another evaluation metric based on the information score \cite{bose2012process}. The current column-wise information score, however, does not quantify the overall information entropy in the whole alignment, which can be useful to evaluate the confidence of individual columns.
	
	Previous studies do not consider the alignment complexity as an individual metric for evaluation \cite{rozinat2007need}\cite{sander1991database}. Trace alignment does not change the type, number or order of the original activities—it only inserts an empty space in traces for which it cannot find a matching activity in a given column. The alignment complexity measures the number of excessive empty spaces that the alignment contains. The optimal alignment has a minimum needed number of empty spaces, while other alignments may have more empty spaces. Although the optimal alignment complexity does not guarantee optimal alignment result, it is better to avoid unnecessary empty spaces, which may cause problems for the column-wise information score, as discussed later. Therefore, it is necessary to consider and quantify alignment complexity as a separate metric.
	
	The reference alignment may often be unavailable because acquiring it is labor intensive, requires domain knowledge, and is subject to human bias. Calculating the optimal trace alignment result may not be practical when the number of traces is large due to the computational complexity \cite{bose2012process}. However, having a reference alignment is necessary to validate an evaluation method and quantify the number of errors in the alignment result. We use an optimum-approaching alignment method called M-COFFEE \cite{wallace2006m} to generate the consensus multi-trace alignment as our reference alignment.
	
	\subsection{Contribution}
	The contributions of this paper are:
	\begin{itemize}
		\item[1] \textbf{Enhancement of misalignment score and information score metrics, and a novel metric of alignment complexity:} We introduced a novel metric of alignment complexity to quantify redundancies in alignment and modified the existing misalignment score and information score metrics to evaluate overall alignment accuracy and confidence. We showed that our methods outperform the previous methods on both synthetic and real-world process logs. We analyzed the influence of evaluation metrics on the overall quality of alignment and proposed a general evaluation procedure that can help identify accurate alignment and optimize the alignment algorithm.
		\item[2] \textbf{Application of trace alignment evaluation to understand medical activities in context:} We validated our proposed evaluation methods on data from a real-world medical process and extracted consensus sequence of activities. We obtained the accurate alignment based on our evaluation methods, analyzed process deviations in activities, and provided medical explanation within a case-study a medical process.
	\end{itemize}
	
	\section{Methodology}\label{sec_2}
	We first describe the trace alignment algorithm, followed by our three alignment metrics: accuracy, confidence, and complexity. We then present our procedure for alignment evaluation.
	\subsection{Alignment Algorithm}
	Trace alignment algorithms include Needleman–Wunsch \cite{needleman1970general}, Smith–Waterman \cite{smith1981identification} and Duration-Aware Trace Alignment \cite{yang2016duration}. Needleman-Wunsch and Duration-Aware Trace Alignment algorithms aim to find globally optimal alignment while Smith–Waterman aims to find locally optimal alignment. Globally optimal alignments align entire activity sequences, from the start until the end, while locally optimal alignment finds the optimal alignment for subprocesses of the process. Duration-Aware Trace Alignment considers the duration of activities, in addition to their sequential order, to generate a globally optimal alignment. We focused on globally optimal alignments because of the following reasons:
	\begin{itemize}
		\item[1] \textbf{The need to model the whole process execution instead of a part of it:} In real-world process logs, finding common patterns between entire traces of process execution is important to extract the workflow. Locally optimal alignments only align segments of the activity sequence, making them not suitable for workflow extraction.
		\item[2] \textbf{Alignment shrinkage due to noise:} Because the locally optimal algorithm aligns the similar subprocesses within process traces, the similar subprocesses are likely to become shorter with increasing number of traces, resulting in alignment shrinkage. In real-world processes, great flexibility and variability of process performance make the common segments very short or nonexistent. Because of the alignment shrinkage problem consensus activities across different process executions cannot be identified. 
	\end{itemize} 
	
	An important issues with globally optimal alignment, is a high time complexity: $O(2^n\ell^n)$, where $n$ is the number of traces and $\ell$ is the average trace length \cite{bose2012process}. This problem is usually addressed by approximating the globally optimal alignment using a heuristic approach or progressive alignment construction \cite{hogeweg1984alignment}. The heuristic approach builds a guide tree connecting each trace so that the alignment iteratively aligns the two closest traces or alignment profiles (intermediate alignment results) in the guide tree until all traces have been aligned. 
	
	An approximation of the globally optimal alignment using a heuristic approach may introduce deviations from the globally optimal alignment, which are called heuristic errors \cite{bose2010trace}. Heuristic errors appear randomly, depending on the guide tree built by the heuristic approach, generating different alignment results with different heuristics. Since the heuristic approach generates no other types of alignment error except for the heuristic errors, we used the number of heuristic errors $N_e$ as an indicator for measuring the alignment accuracy.
	
	\subsection{Improved Metric: Accuracy}
	The alignment accuracy metric measures the number of activities correctly aligned and is directly correlated to the $N_e$. Alignment accuracy evaluation methods include sum-of-pairs score (SPS), column score (CS), and misalignment score.
	\subsubsection{Sum-of-pairs Score (SPS)}
	\paragraph{Reference-based sum-of-pairs score}The reference-based sum-of-pairs score, sometimes called quality score (Q score) \cite{sauder2000large}, is the total number of correctly aligned activity pairs in the alignment result divided by the number of all aligned activity pairs in the reference alignment. Correct alignment means that the aligned activity pairs in alignment result are also aligned in the reference alignment. The reference-based sum-of-pairs score has two derivatives: developer score and modeler score \cite{sauder2000large}. The derivatives only differ in terms of choosing the reference alignment and are consistent with each other in the same process log.
	\paragraph{Reference-free sum-of-pairs score}Reference-free sum-of-pairs score assumes that aligning activities of the same type is correct (``match''), aligning activities to different type activities is incorrect (``mismatch'') and aligning activities to the empty space is acceptable (``gap'') \cite{setubal1997introduction}. Mismatches are given a penalty while gap penalties vary across different scoring schemes. The reference-free sum-of-pairs score measures the similarity between traces within the alignment. The higher similarity indicates a better alignment quality. We adopted a commonly used scoring scheme: match = 1, mismatch = -1 and gap = 0 \cite{setubal1997introduction}.
	
	\subsubsection{Column Score (CS)}
	The column score is defined as the number of correctly aligned columns divided by the total number of columns in the alignment \cite{thompson1999comprehensive}. Here, ``correctly aligned'' means that the types and numbers of activities in an alignment column are exactly the same as that in the corresponding reference column.
	
	\begin{figure}[!t]
		\begin{center}
			\includegraphics[width=0.5\columnwidth]{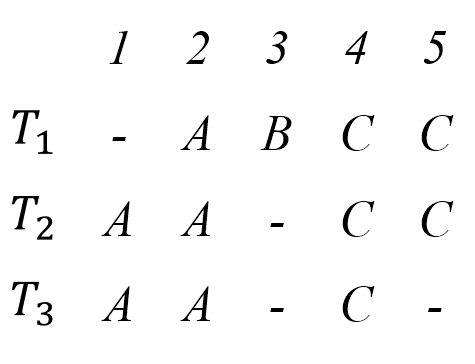}
			\\
			\textbf{(a)}
		\end{center}
		\begin{center}
			\includegraphics[width=\columnwidth]{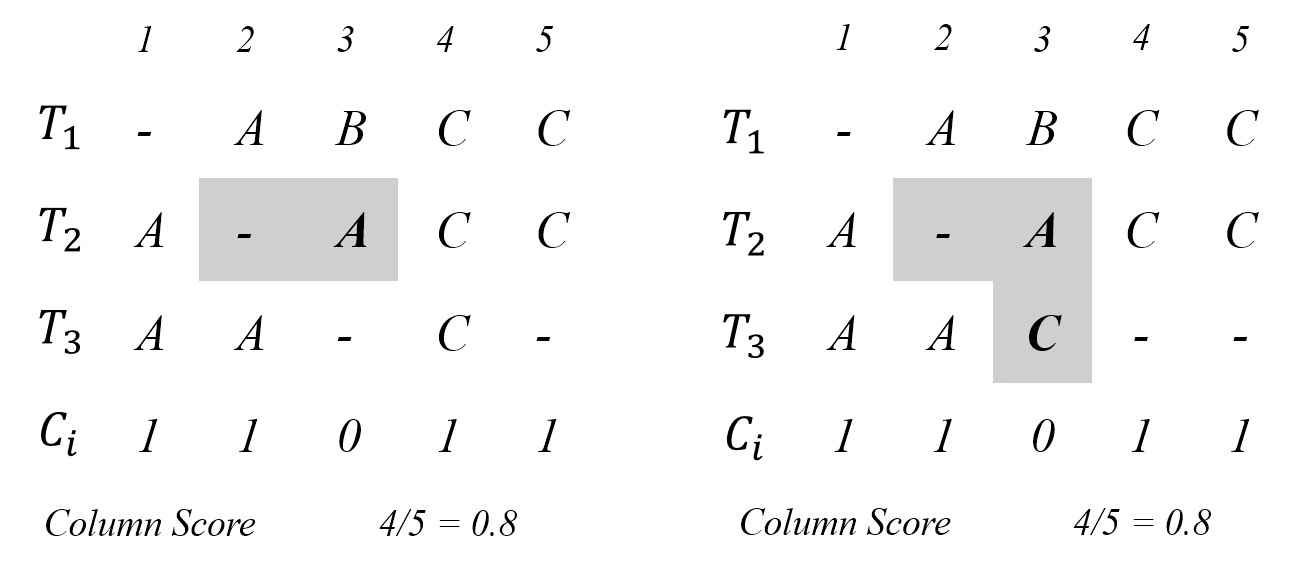}
			\\
			\textbf{(b)}\hspace{1.5in}
			\textbf{(c)}
		\end{center}
		\caption{\textbf{(a)} Reference alignment \textbf{(b)} Alignment with 1 misalignment \textbf{(c)} Alignment with 2 misalignments (misaligned activities are shaded in gray)}
		\label{fig_CS}
	\end{figure}
	
	Column score is highly sensitive to misaligned activities. Any activity in an alignment column different from that in the reference will make the column counted as misaligned. It does not distinguish the number of misaligned activities within the same column (\figurename{\ref{fig_CS}}), making it a coarse measure of alignment accuracy.
		
	\subsubsection{Misalignment Score}
	Alignment accuracy can also be quantified by the misalignment score. In a trace, consecutive activities with a specific order form a certain pattern, and the misalignment score considers the sequential order of these activities in the pattern. For a specific pattern, misalignment score measures the similarity between traces by checking if the patterns are aligned in the same columns. If not, misalignment score measures the distance between each column the patterns are in and sums the distances. Misalignment scores of $N$ traces form a matrix containing pairwise misalignment scores $ms_{ij}$ \cite{bose2012process}:
	\begin{equation}
		\label{eq_1}
		{ms}_{ij} = ms_{ji} = \sum_{k=1}^{K}\sum_{r\in {IN(p,k,\overline{T_i},\overline{T_j})}}\left|r-k\right|+\delta(k)
	\end{equation}
	where $K$ denotes the larger number of pattern repetitions between trace $\overline{T_i}$, $\overline{T_j}$; ${IN(p,k,\overline{T_i},\overline{T_j})}$ is the mapping set of pattern instances in $\overline{T_j}$ and $\overline{T_i}$; $\delta(k)$ is the score of $k^{th}$ pattern instance in $\overline{T_i}$ (if the $k^{th}$ instance is aligned to a gap or other activity that is not in the pattern, then $\delta(k)=1$; otherwise $\delta(k)=0$) \cite{bose2012process}. 
	
	Note that the pattern considered in the misalignment score does not include the gap symbol ``-''. The higher misalignment score indicates less similarity between traces, meaning that the traces are considered misaligned. The misalignment score adds up all pairwise misalignment scores:
	\begin{equation}
		\label{eq_2}
		MS = \sum_{i,j}ms_{ij}
	\end{equation}
	
	Because the misalignment score measures the degree of misalignment with respect to a certain pattern in the alignment, the crucial parameter is the pattern. However, the choice of the pattern has not been established. The problem exists in situations where a single pattern may not reflect the misalignments in the whole alignment. In addition, the existing misalignment score metric does not consider the pattern’s frequency, which also influences the quality. To achieve a more comprehensive view of misalignments, we modified the method for selecting the pattern and defined an overall misalignment score based on the patterns in the alignment.
	
	\subsubsection{A Novel Overall Misalignment Score (OMS)}
\begin{figure}[!b]
		\begin{center}
			\includegraphics[width=\columnwidth]{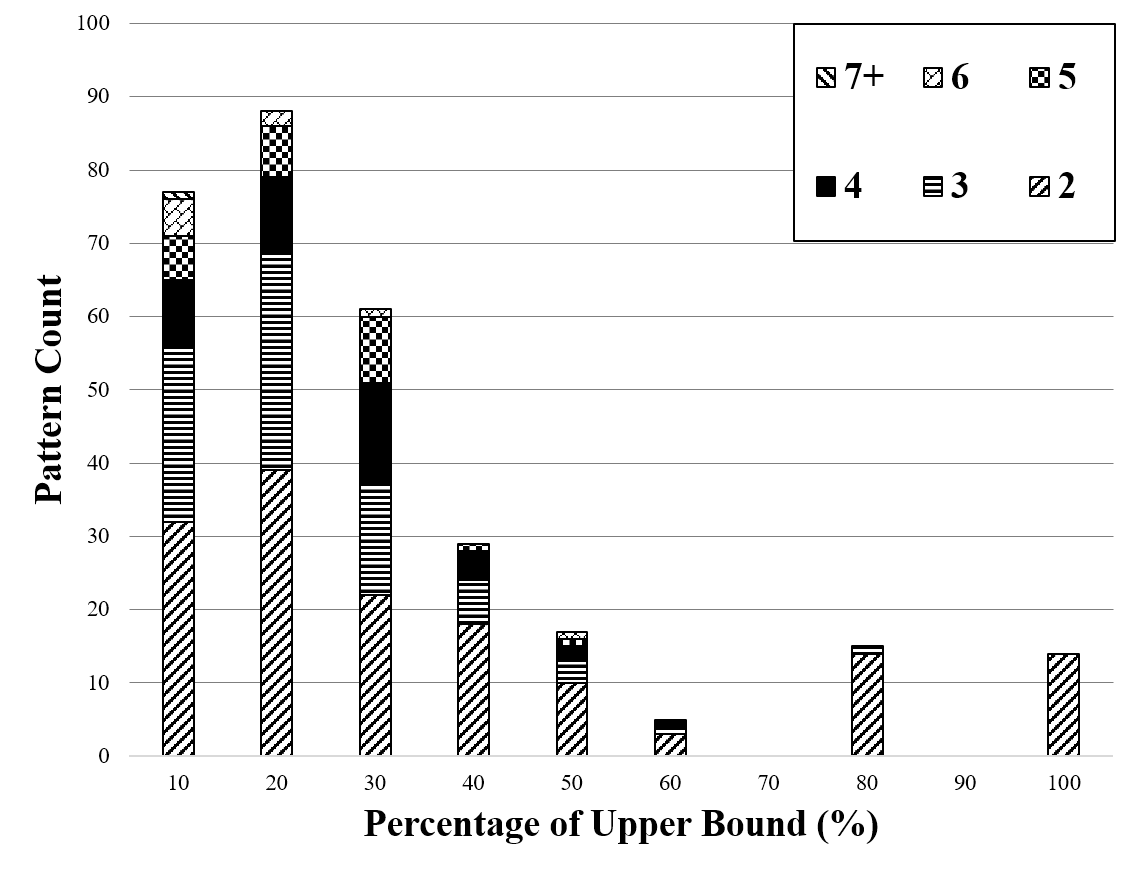}
		\end{center}
		\caption{Pattern count and length distribution over frequency of a synthetic log data. The synthetic logs are generated using PLG \cite{burattin2010plg} based on 5 simplified medical process models established by medical experts. The X-axis shows the percentage interval of each pattern's frequency over the upper bound ($f_M$), Y-axis is the count of patterns. Patterns with different length are marked in differently bars and corresponding length is in the legend.}
		\label{fig_distribution}
	\end{figure}

	Frequency is one of the most intuitive ways to measure representativeness. As mentioned, misalignment score depends on the pattern chosen, which is supposed to be representative for the whole alignment. Given that patterns vary in length and have a different contribution to the alignment accuracy: longer patterns tend to be much rarer than the shorter patterns and longer patterns are more likely to be misaligned since they require aligning more activities. The pattern's length and frequency have a characteristic distribution (\figurename{\ref{fig_distribution}}): the longer patterns (with a length greater than 5) are much less frequent than the shorter patterns. Hence, when measuring the misalignments of patterns in the whole alignment, distinguishing the patterns based on their frequencies is intuitively more precise to depict the distribution of misalignments than using the most frequent pattern only.
	
	We chose the patterns that occur more frequently than a threshold $T_f$ for calculating misalignment score. Each pattern's misalignment score is assigned a weight based on the ratio of pattern’s frequency over the frequency of the most frequent pattern. In this weighting method, a pattern of a higher frequency has a larger influence in the misalignment score. Our overall misalignment score ($OMS$) is:
	\begin{equation}
		\label{eq_OMS}
		OMS = \frac{1}{N}\sum_{f_p>T_f}{MS}_{p}\times\frac{f_p}{f_M}
	\end{equation}
	
	where $f_p$ is the occurrence of the pattern $p$, ${MS}_p$ is the original misalignment score for pattern $p$, $f_M$ is the maximum occurrence and $N$ is the number of patterns. This overall misalignment score considers all eligible patterns, and the alignment accuracy mainly depends on shorter but frequent patterns, instead of longer but infrequent ones.

	\begin{table}[!t]
		\renewcommand{\tabularxcolumn}[1]{>{\small}m{#1}}
		\newcolumntype{Y}{>{\centering\arraybackslash}X}
		\renewcommand{\arraystretch}{1.3}
		\caption{Overall misalignment score's correlation to $N_e$ with different frequency threshold $T_f$ settings$^\dagger$}
		\label{table_freqThres}
		\centering
		\begin{tabularx}{\columnwidth}{|>{\centering}m{.03\textwidth}|>{\centering}m{.03\textwidth}|>{\centering}m{.03\textwidth}*{2}{|Y}|}
			\hline
			Log &$f_M$&$T_f$ &$T_f/f_M$ \%&$OMS$'s correlation to $N_e$\\
			\hline
			\hline
			\multirow{5}{*}{1}
			&\multirow{5}{*}{10}
			&2&20.00&0.7048\\\cline{3-5}
			&&\textbf{4}&\textbf{40.00}&\textbf{0.9834}\\\cline{3-5}
			&&6&60.00&0.9650\\\cline{3-5}
			&&8&80.00&0.8934\\\cline{3-5}
			&&10&100.00&0.8695\\\cline{3-5}
			\hline
			\multirow{5}{*}{2}
			&\multirow{5}{*}{27}
			&5&18.52&0.8380\\\cline{3-5}
			&&\textbf{11}&\textbf{40.74}&\textbf{0.9539}\\\cline{3-5}
			&&16&59.26&0.9023\\\cline{3-5}
			&&22&81.48&0.7375\\\cline{3-5}
			&&27&100.00&0.6193\\\cline{3-5}
			\hline
			\multirow{5}{*}{3}
			&\multirow{5}{*}{32}
			&7&21.88&0.6371\\\cline{3-5}
			&&\textbf{14}&\textbf{43.75}&\textbf{0.9778}\\\cline{3-5}
			&&20&62.50&0.8762\\\cline{3-5}
			&&26&81.25&0.8506\\\cline{3-5}
			&&32&100.00&0.7790\\\cline{3-5}
			\hline
			\multirow{5}{*}{4}
			&\multirow{5}{*}{57}
			&14&24.56&0.7159\\\cline{3-5}
			&&26&45.61&0.7632\\\cline{3-5}
			&&\textbf{36}&\textbf{63.16}&\textbf{0.7711}\\\cline{3-5}
			&&47&82.46&0.7172\\\cline{3-5}
			&&57&100.00&0.6850\\\cline{3-5}
			\hline
			\multirow{5}{*}{5}
			&\multirow{5}{*}{164}
			&33&20.12&0.7946\\\cline{3-5}
			&&\textbf{65}&\textbf{39.63}&\textbf{0.8768}\\\cline{3-5}
			&&98&59.76&0.8319\\\cline{3-5}
			&&131&79.88&0.8420\\\cline{3-5}
			&&164&100.00&0.8152\\\cline{3-5}
			\hline
			\multicolumn{5}{@{}l}{$\dagger$\footnotesize{The setting of highest correlation for each log is in \textbf{bold}}}
		\end{tabularx}
	\end{table}
	
	\begin{table}[!b]
		\renewcommand{\tabularxcolumn}[1]{>{\small}m{#1}}
		\newcolumntype{Y}{>{\centering\arraybackslash}X}
		\renewcommand{\arraystretch}{1.3}
		\caption{Correlation between misalignment scores and the number of heuristic errors $N_e$ in different logs}
		\label{table_OverallMSCorr}
		\centering
		\begin{tabularx}{\columnwidth}{*{3}{|Y}|}
			\hline
			Log &OMS &MS$^\dagger$\\
			\hline
			\hline
			1 & 0.9834&0.9702\\
			\hline
			2 & 0.9539&0.9025\\
			\hline
			3 & 0.9778&0.9457\\
			\hline
			4 & 0.7632&0.6903\\
			\hline
			5 & 0.8768&0.7308\\
			\hline
			\multicolumn{3}{@{}l}{$\dagger$\footnotesize{Based on the most frequent pattern in the alignment}}		
		\end{tabularx}
	\end{table}
	Since frequency threshold $T_f$ affects the evaluation method's ability to evaluate misalignments, we would like to maximize the correlation between the overall misalignment score $OMS$ and the number of errors $N_e$ by choosing an appropriate $T_f$. We determined the correlation between $OMS$ and $N_e$ with different $T_f$ settings on different synthetic logs (\tablename{\ref{table_freqThres}}), and found that $T_f \approx 40\%\times f_M$ makes $OMS$ the highest correlation to $N_e$ except for the $4^{th}$ log (\tablename{\ref{table_freqThres}}) (although $T_f = 45.61\%$ of $f_M$ in the $4^{th}$ log comes close to the highest correlation of \textbf{0.7711}). Thus, we recommend using the frequency threshold $T_f$ as 40\% of $f_M$ to choose patterns for the overall misalignment score calculation.
	
	To compare our $OMS$ metric \ref{eq_OMS} to the original misalignment score $MS$ \ref{eq_2}, we analyzed the correlation of the $OMS$  and $MS$ to $N_e$ on the same synthetic data sets (\tablename{\ref{table_OverallMSCorr}}). We generated 30 alignments with different $N_e$ of each log. The results showed that $OMS$ has a higher correlation to $N_e$, indicating our $OMS$ performed better in evaluating the overall alignment accuracy than the original misalignment score.
	
	The time complexity of calculating the overall misalignment score is $O(N{\ell_{max}}^3+N^2{\ell_{max}})$, where extracting patterns takes $O(N{\ell_{max}}^3)$ time, and calculating misalignment score for each pattern takes $O(N^2{\ell_{max}})$ time; $N$ is the number of traces and ${\ell_{max}}$ is the longest trace length. 
	\subsection{Improved Metric: Confidence}
	\begin{figure}[!b]
		\begin{center}
			\includegraphics[width=3in]{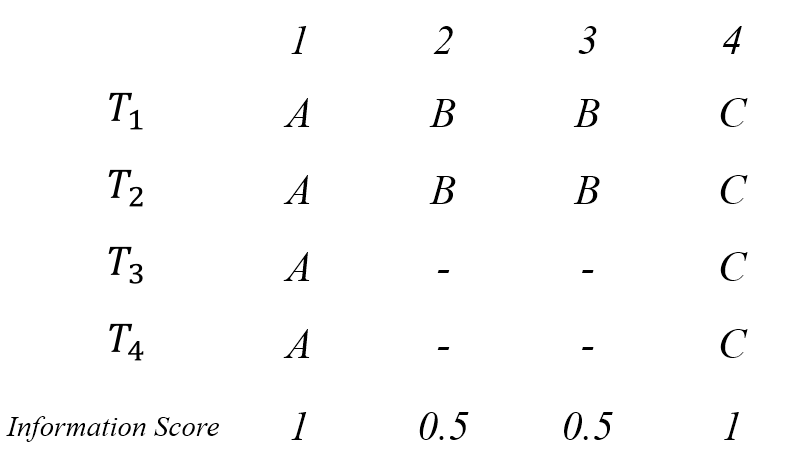}
			\\
			\textbf{(a)}
		\end{center}
		\begin{center}
			\includegraphics[width=3in]{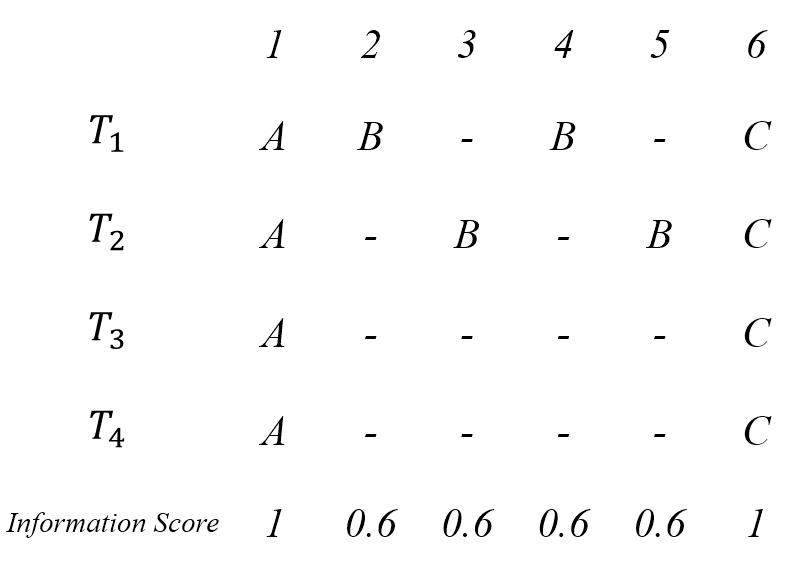}
			\\
			\textbf{(b)}
		\end{center}
		\caption{\textbf{(a)} Reference alignment \textbf{(b)} Preferred alignment by information score}
		\label{fig_information score}
	\end{figure}
	\subsubsection{Information Score}
	As mentioned, a single column's confidence in the alignment result is measured by information score. Information score is a quantification method based on information entropy considering activity types and frequencies. Each type of activity has a frequency and its information entropy can be calculated based on its frequency \cite{bose2012process}:
	\begin{equation}
		\label{eq_4}
		IS = 1-\frac{E}{E_{max}}
	\end{equation}
	\begin{equation}
		\label{eq_5}
		E = \sum_{i=1}^{n}-P_i\log_2{P_i}
	\end{equation}
	where $IS$ is the column’s information score, E is the column’s information entropy and $P_i$ is the occurrence frequency of each type’s activity in the column; $E_{max}$ is the maximum entropy of the whole alignment which equals $-\log_2{(N+1)}$, $N$ is the number of activity types ($+1$ for the gap activity). 
	
	The purpose of trace alignment is to discover consensus activities and to detect deviations, and a high alignment confidence helps in achieving the purpose. According to (\ref{eq_4}) and (\ref{eq_5}), if all types of activities (including gap) occur in the column with same frequency, the information score will reach the minimum value of 0, meaning low confidence of the column; if the column is filled with only one type of activity, the information score will reach the maximum value of 1, indicating high confidence.
	
	Problems arise when considering information score for columns only: To obtain higher information score, the alignment tends to \textbf{(1)} split columns if there are two or more activities in the column and the gaps cannot be replaced by activities in other columns; or \textbf{(2)} align activities incorrectly in one column if the gaps in this column can be replaced by shifting other activities. 
	
	In situation \textbf{(1)}, the alignment tends to split activities that are already aligned. In this case, each column’s information score will be higher but the alignment is unreasonably longer. If the target alignment \textbf{(a)} has 50\% of activity B in column 2 and 3, the preferred alignment of information score \textbf{(b)} will align each B in individual columns 2, 3, 4 and 5 (\figurename{\ref{fig_information score}}). In situation \textbf{(2)}, the alignment algorithm tends to shorten the alignment by aligning incorrect activities in a column (\figurename{\ref{fig_information score}}). Thus, if information score is applied to evaluate alignment algorithm column-by-column, the algorithm tends to split or merge columns incorrectly.
	
	\subsubsection{Proposed Overall Information Score (OIS)}
	To address the problems with column-wise information score, we added a constraint on the cumulative information entropy for all columns:
	\begin{equation}
		\label{eq_MIE}
		E = \sum_{j=1}^{L}{\sum_{i=1}^{n}-P_{ji}\log_2{P_{ji}}}\times{\frac{1}{L}}
	\end{equation}
	where $j$ is the column index, $L$ is the alignment length. Then we propose our overall information score ($OIS$) as:
	\begin{equation}
		\label{eq_OIS}
		OIS = 1-\frac{\sum_{j=1}^{L}{\sum_{i=1}^{n}-P_{ji}\log_2{P_{ji}}}}{E_{max}\times L}
	\end{equation}
	
	With the constraint, overall information score will not keep increasing monotonously with the increasing of columns, if the gain in the information entropy of the splitting column does not exceed the loss of average information entropy of the whole alignment, the splitting column operation is considered unnecessary and will not be proceeded.
	
	The overall information score reduces the unnecessary column-splitting problem of evaluating the alignment confidence using column-wise information score and reflects the overall confidence in the whole alignment.
	\begin{figure}[!b]
		\begin{center}
			\includegraphics[width=\columnwidth]{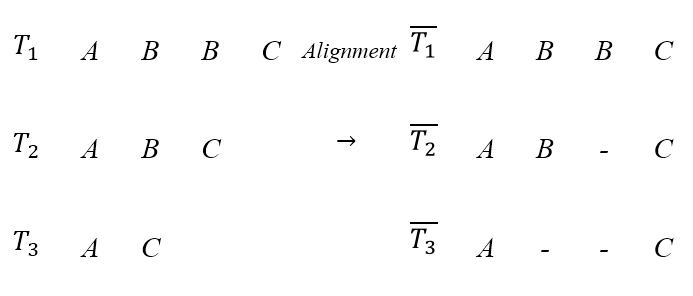}
			\\
			\textbf{(a)}
		\end{center}
		\begin{center}
			\includegraphics[width=\columnwidth]{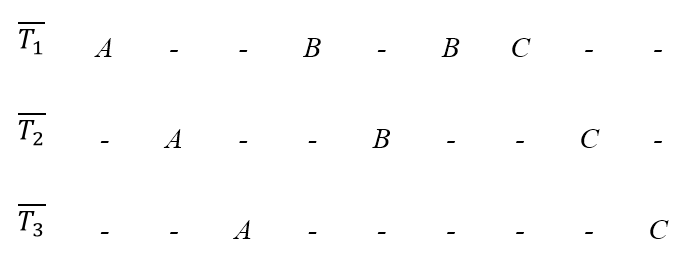}
			\\
			\textbf{(b)}
		\end{center}
		\caption{\textbf{(a)} Absolute lower bound (0.25) and \textbf{(b)} Absolute upper bound (0.67) of gaps number (18 of "-") $M=9$, $N=3$ and $L_{min} = 4$}
		\label{fig_complexity}
	\end{figure}
	
	\begin{table*}[!h]
		\renewcommand{\arraystretch}{1.3}
		\caption{Attributes considered for evaluating alignment and their monotony}
		\label{table_attributes}
		\centering
		\begin{tabular}{|c|c|c|c|c|}
			\hline
			Methods &Metric&Attributes Considered&Score Range&Monotony\\
			\hline
			\hline
			Reference-free sum-of-pairs score&Accuracy&Activities type & $[0,?^\dagger)$&$\nearrow$ when $N_e\searrow$\\
			\hline
			Reference-based sum-of-pairs score&Accuracy&Activities type &$[0,1]$&$\longrightarrow 1$ when $N_e\searrow$\\
			\hline
			Column score &Accuracy&Activities type & $[0,1]$&$\longrightarrow 1$ when $N_e\searrow$\\
			\hline
			Modified Misalignment score&Accuracy&Patterns type, frequency &$[0,+\infty)$&$\nearrow$ when $N_e\nearrow$\\
			\hline
			Modified Information score&Confidence&Activities type, frequency &$[0,1]$&$\longrightarrow 1$ when information amount $\nearrow$\\
			\hline
			Alignment complexity&Complexity&Activities frequency, alignment length &$[0,1]$&$\longrightarrow 1$ when alignment length $\nearrow$\\
			\hline
			\multicolumn{5}{@{}l}{$\dagger$\footnotesize{The upper bound for reference-free sum-of-pairs score depends on the data set, and does not have a fixed value}}		
		\end{tabular}
	\end{table*}
	\subsection{Novel Metric: Complexity}
	In current evaluation methods, the length is not considered as an individual metric \cite{de2012multi}; however, based on the previous discussion, we found that alignment length has a significant influence on alignment evaluation methods, including column-wise information score and column score. Alignment length also reflects the computational complexity needed to perform alignment: longer alignments contain more places for filling activities, resulting in higher computational complexity. In this section, we propose our new metric of alignment complexity based on the alignment length and the number of activities.
	
	The alignment complexity can be calculated by the percentage of gaps: some gaps in the alignment are unnecessary, and so the percentage of gaps reflects the degree of redundancy. Yet not all the gaps are unnecessary since aligning activities requires gaps to place activities that cannot be aligned. 
	
	Note that there is a minimum number of gaps required to accomplish an alignment. Since traces in the alignment are of the same length, the differences in original traces’ length will be filled by gaps. The number of gaps required to fill the traces to the maximum length is the lower bound of alignment complexity. For example, for three traces $T_1$, $T_2$ and $T_3$ (\figurename{\ref{fig_complexity}}), the minimum number of gaps needed is 3. In this example, no extra gaps are added with all columns correctly aligned (\figurename{\ref{fig_complexity}} \textbf{(a)}). 
	
	There is also an upper bound of alignment complexity, which occurs when every column in the alignment contains only one activity from the original trace log (\figurename{\ref{fig_complexity}} \textbf{(b)}). Thus, the alignment complexity $P$ of an alignment can be written as:
	
	\begin{equation}
		\label{eq_6}
		P=1-\frac{M}{N\times L}
	\end{equation}
	
	\begin{equation}
		\label{eq_7}
		1-\frac{M}{(N\times L_{min})} \leq P \leq 1-\frac{1}{N}
	\end{equation}
	$M$ is the number of activities in the original trace log, $N$ is the number of traces, $L$ is the alignment length and $L_{min}$ is the shortest length of alignment, which is also the longest original trace's length.
	
	The lower alignment complexity means alignment has less redundancy. However, the optimal alignment does not guarantee the lowest alignment complexity. Due to this limitation of alignment’s complexity, it should be considered with the lowest priority when combined with other methods.
	
	\subsection{General Evaluation Procedure}
	With the different trace alignment quality metrics, the attributes considered in each evaluation method and their monotony are summarized in \tablename{\ref{table_attributes}}. When evaluating an alignment with a reference alignment, higher reference-based sum-of-pairs score means the alignment is closer to the reference alignment, and lower misalignment score indicates the alignment has a lower degree of misalignment in the alignment's patterns; a higher information score means more confidence in the common patterns and deviations found in the alignment, and a lower alignment complexity is expected for lower redundancy. The reference-free evaluation does not use the reference-based sum-of-pairs score or column score, while the other evaluation methods are the same with the reference-based evaluation.
	
	\begin{table*}[!h]
		\renewcommand{\tabularxcolumn}[1]{>{\small}m{#1}}
		\newcolumntype{Y}{>{\centering\arraybackslash}X}
		\renewcommand{\arraystretch}{1.3}
		\caption{Evaluation methods correlation with the number of heuristic error in alignments$^\dagger$}
		\label{table_evaluationresult}
		\newcommand{\colwidth}{0.53in}
		\centering
		
		\begin{tabularx}{\textwidth}{|
				>{\centering}m{.14\textwidth}|>{\centering}m{.06\textwidth}
				*{8}{|Y} |}
			\hline
			Data set&$N_e$&Ref-free SPS&Ref-based SPS&MS$^\ddagger$&OMS&OIS&Column Score&Alignment Complexity\\
			\hline
			\hline
			\multirow{13}{*}{Primary Survey}	
			&61&1478&0.4833&5.3097&5.0424&0.5793&0.3902&0.8203	\\\cline{2-9}
			&52&1422&0.4493&5.2909&5.0242&0.5837&0.4186&0.8287	\\\cline{2-9}
			&43&1669&0.5769&5.2776&4.9818&0.5902&0.4103&0.8112	\\\cline{2-9}
			&42&1615&0.5459&5.2685&4.9758&0.5949&0.4146&0.8204	\\\cline{2-9}
			&36&1684&0.5803&5.2358&4.9455&0.5794&0.4324&0.8010	\\\cline{2-9}
			&25&1810&0.6443&4.7818&4.4182&0.5676&0.4688&0.7699	\\\cline{2-9}
			&20&1932&0.7011&4.7076&4.3576&0.5789&0.5161&\textbf{0.7625}	\\\cline{2-9}
			&19&1814&0.6298&4.0961&3.9576&0.5729&0.3939&0.7769	\\\cline{2-9}
			&13&1689&0.5798&3.7697&3.6727&0.5663&0.2857&0.7896	\\\cline{2-9}
			&5&1782&0.5958&\textbf{3.7494}&\textbf{3.6000}&0.5735&0.3235&0.7834	\\\cline{2-9}
			&0 (\textbf{Ref})&\textbf{2297}&\textbf{1.0000}&4.5924&4.3091&\textbf{0.6545}&\textbf{1.0000}&0.7834	\\\hline\hline
			Correlation to $N_e$&/&-0.8188&-0.7346&0.8187&\textbf{0.8547}&-0.2594&-0.3805&0.7875	\\\hline
			\multicolumn{9}{@{}l}{$\dagger$\footnotesize{The best results are marked in \textbf{bold}, and the alignment with 0 $N_e$ is used as the reference alignment}}\\
			\multicolumn{9}{@{}l}{$\ddagger$\footnotesize{The pattern chosen for original misalignment score is the most frequent pattern in the alignment regardless of the pattern's length}}		
		\end{tabularx}
	\end{table*}
	
	Currently, there is no general procedure for trace alignment evaluation \cite{adriansyah2012alignment}. To evaluate the overall alignment quality, evaluation metrics should be standardized. Alignment accuracy, confidence, and our proposed complexity metrics are used to evaluate the overall quality of the alignment results. Though measured independently, they are not strictly orthogonal or independent of each other because:
	\begin{itemize}
		\item[1] The attributes taken into consideration are not exclusive or inclusive. Accuracy, confidence, and complexity involve different attributes including activities types, frequency, patterns types and alignment length. These attributes are not orthogonal in the feature space, though they have a certain degree of overlap, i.e., patterns frequency is correlated to the frequency of each activity in the pattern, which makes overall misalignment score partially related to overall information score. 
		
		\item[2] Though some correlation exists between the evaluation metrics, the correlation cannot be quantified generally since it is data-specific. The correlation between overall misalignment score and information score depends on the frequent activities and patterns in the data; the activities and patterns differ between data sets and may result in correlation coefficients ranging from -1 to 1.  
	\end{itemize}
	
	Although they may overlap, each evaluation metric describes a unique aspect of the alignment, and should be considered in this order:
	
	Firstly, alignment accuracy is the most important metric, as it is directly related to the number of alignment errors. Since the column score only provides a coarse evaluation and is sensitive to deviations, it is considered less significant when other accuracy methods are available.
	
	Then the alignment confidence is measured by overall information score. A high overall information score is expected for alignment with strong confidence in patterns and deviations extracted. The overall information score will not increase if the alignment tries to split aligned columns. 
	
	Alignment complexity is considered with the lowest priority because it does not contribute to the consensus activities found or the amount of information in the alignment. However, alignment complexity can be used to avoid unnecessary computational complexity and reduce redundancies in the alignment.
	
	\section{Experiment and analysis}\label{sec_3}
	\subsection{Experiment Design}
	Needleman–Wunsch and Duration-Aware Trace Alignment algorithms generate different alignments. A reference alignment was set with $N_e = 0$ based on the consensus multi-trace alignment using M-COFFEE. We detected the number of heuristic errors $N_e$. The same log using different parameters generated alignments with different $N_e$. Using these, we showed the correlation between evaluation results and the number of heuristic errors. The high absolute value of correlation indicates the evaluation method truly reflects the alignment accuracy. Though alignment confidence and complexity are not designed for measuring alignment accuracy, their correlation was computed to check if they have overlap with alignment accuracy evaluation methods.
	
	The log in this experiment includes 33 individual trauma resuscitation cases with 247 activities of 14 types. The data set was collected from August to December 2014. The log was coded by medical experts reviewing videos of trauma resuscitation. The use of this data set and its related research has been approved by the IRB of the Children’s National Medical Center.
	
	\subsection{Results}
	\subsubsection{Alignment Evaluation}
	The evaluation results show the correlation between different evaluation methods and the number of error $N_e$ (\tablename{\ref{table_evaluationresult}}). The overall misalignment score has the highest correlation to the number of heuristic errors, while reference-free sum-of-pairs score, reference-based sum-of-pairs score, misalignment score and alignment complexity also achieve relatively high correlation. This shows our overall misalignment score better reflects the alignment accuracy than the original misalignment score and sum-of-pairs score. Our proposed alignment complexity also has a high correlation to the $N_e$ here because correcting heuristic error often results in shorter alignments, and decreases the alignment complexity. However, this is not always guaranteed. If the heuristic error is not the only activity in the column, correcting the error may not decrease the alignment length.
	
	The reference alignment performs best in reference-free sum-of-pairs score, reference-free sum-of-pairs score, overall information score, and column score, while the alignment with 5 heuristic errors performs best in misalignment score and overall misalignment score, and the alignment with 20 heuristic errrs performs best in alignment complexity. The alignment with 5 heuristic errors aligns frequent patterns correctly, but misaligned some patterns with lower frequency; the reference alignment aligns most long patterns correctly except some patterns of length 2. Since the overall misalignment score considers pattern frequency, and shorter patterns are probably more frequent, the reference alignment here does not perform best in the misalignment score. The alignment with 20 heuristic errors has a short alignment length, but the low complexity comes at the price of aligning activities evenly among columns instead of aligning them to the most other activities. This confirms our assumption that alignment complexity should not be a prioritized metric. However, lower alignment complexity is still preferable when other metrics are the same, since it reduces the computation complexity and redundancy in the alignment.
	
	\begin{figure}[!t]
		\begin{center}
			\includegraphics[width=0.9\columnwidth]{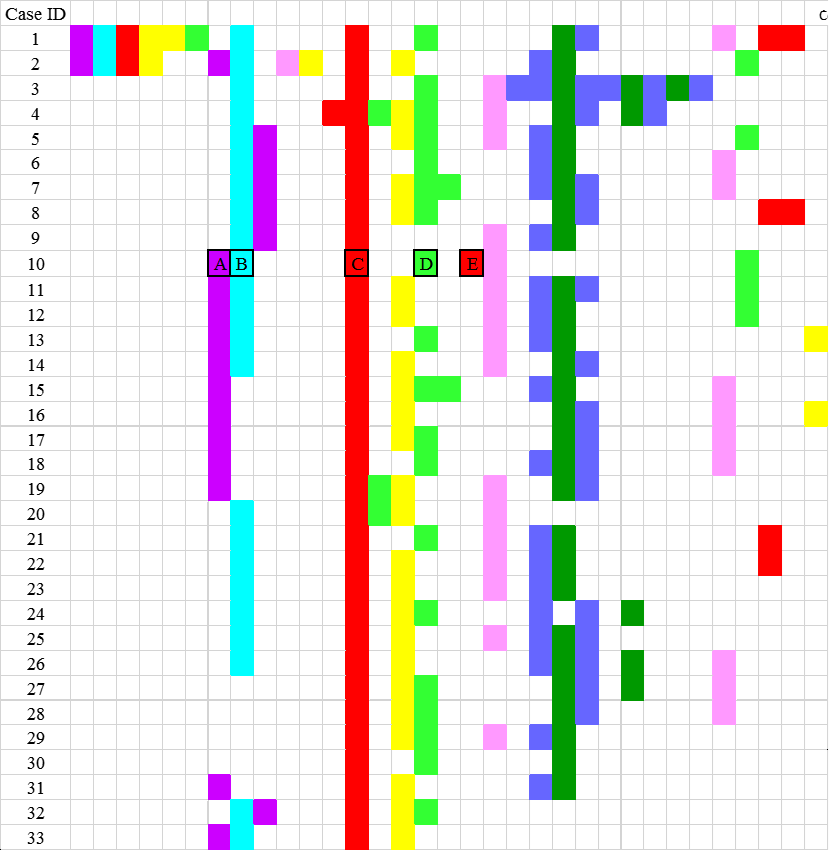}
			\\
			\textbf{(a)}
		\end{center}
		\begin{center}
			\includegraphics[width=0.9\columnwidth]{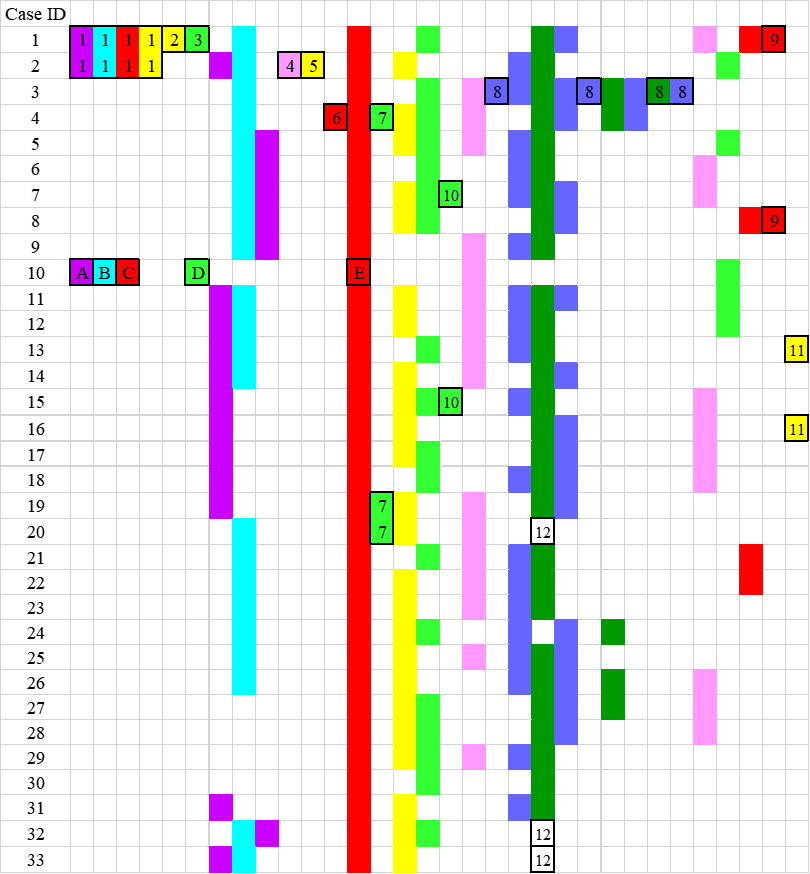}
			\\
			\textbf{(b)}
		\end{center}
		\begin{center}
			\includegraphics[width=0.7\columnwidth]{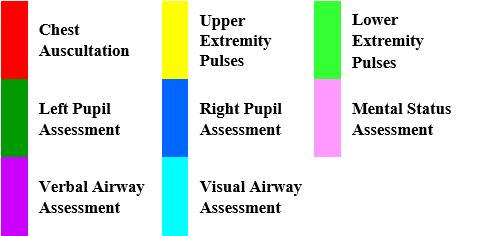}
			\\
		\end{center}
		\caption{\textbf{(a)} Alignment with 5 heuristic errors labeled in A, B, C, D and E \textbf{(b)}Reference alignment in \tablename{\ref{table_evaluationresult}}, deviations labeled from 1 to 12. See digital version for color and higher resolution}
		\label{fig_NW_Ref}
	\end{figure}
	
	The evaluation methods indicate the alignment with 5 heuristic errors has the best performance in the overall misalignment score and relatively good performance in the sum-of-pairs score and alignment complexity. The alignment with 5 heuristic errors \figurename{\ref{fig_NW_Ref}}\textbf{(a)} have aligned activities $A, B, C, D$ to the majority activities, while activity $E$ is aligned in a single column as a deviation. The alignment with 5 heuristic errors has lower overall misalignment score than the reference alignment \figurename{\ref{fig_NW_Ref}}\textbf{(b)}, which aligns activities $A, B, C, D$ to the deviation activities in cases 1 and 2, and aligns activity $E$ to the majority. The reference alignment has a better sum-of-pairs score, overall information score and alignment complexity.
	
	\subsubsection{Medical Explanation}
	The reference alignment attempts to align activities $A, B, C, D$ to those activities in cases 1 and 2. Cases 1 and 2 are similar because the patient arrived before the complete medical team could assemble. In both cases, someone other than the examining provider began the assessment, and upon arrival, the assigned examining provider started the exam over from the beginning, producing the repeated activity sequences shown in cases 1 and 2. In 10 the full medical team had assembled before the patient’s arrival, and the examining provider was the only person to conduct the examination. Although this alignment produced a unique deviation of $E$ due to a repeated chest auscultation, a review of the data showed that the rest of the activities were more consistent with the consensus sequence derived from this algorithm. Above all, the alignment with 5 heuristic errors is more clinically similar to the cases which it is aligned. 
	
	Our overall misalignment score evaluation matches the feedback from the medical team because the overall misalignment score considers the performance of all patterns in the alignment. Though the reference alignment has a better sum-of-pairs score, confidence and complexity, it comes at the price of misaligning the patterns $<C, D>$ and $<D, E>$. In real-life situations, patterns are usually considered with priority since the alignment is context-based. With the high correlation between overall misalignment score and number of heuristic errors, we can use the overall misalignment score as the principle alignment accuracy evaluation method.
	
	Besides the 5 heuristic errors in the alignment, we also analyzed the activities considered to be deviations in both of the alignments. These activities are labeled 1 to 12 and marked in black boxes \figurename{\ref{fig_NW_Ref}}\textbf{(b)}. 
	
	For deviations $1, 2, 3, 5, 6, 9$, these activities were performed by different roles during the trauma resuscitation. Since the alignment algorithm does not consider role information and aligns activities based on activity type and order only, these activities are aligned in single columns with no other optimal places.
	
	Activity $4$ was a deviation since the person performing this activity took action earlier than expected, making this activity out of order; activities $7$ should be generally performed after Upper Extremity Pulses Assessment because Lower Extremity Pulses Assessment is checked at patient's feet, and the examining provider typically proceeds from the head to the feet, yet occasionally Lower Extremity Pulses Assessment can be performed before; activities $8$ were deviations because the patient was not cooperative and medical team had to perform pupil assessment multiple times; activities $10$ were minor deviations due to the back and forth assessments performed by the medical team; activity $11$ in case 13 was skipped at first and then the medical team performed it after pupil assessment, and activity $11$ in case 16 was an unexpected repetition; activities $12$ were performed in later phases and thus not included in the primary survey phase, making the cases 20, 32 and 33 skip the pupil assessments, and these disappearances of pupil checks were considered deviations during the trauma resuscitation.
	
	The good quality alignments are able to identify real-life deviations. Though some deviations like $1, 2, 3, 5, 6, 9$ in 
	\figurename{\ref{fig_NW_Ref}}\textbf{(b)} are false alarms due to unconsidered role information, the overall evaluation methods still help to generate the alignment with a minimal number of heuristic errors.
	
	\section{Conclusion}\label{sec_4}
	In this paper, we analyzed the previous alignment evaluation methods and proposed our modification to make them suitable for evaluating overall alignment quality. We discussed the limitations of previous evaluation methods through experiments on synthetic process log data, showing that column score is not appropriate for precise accuracy measurement. Experimental results showed our overall misalignment score and overall information score perform better in evaluating alignment quality than the previous ones. We verified our methods on a real-life medical process, showing our evaluation methods help identify deviations in the process. With the overall alignment quality evaluation methods, the alignment algorithm can be further optimized.
	
	\section*{Acknowledgment}
	This research was supported by the National Library of Medicine of the National Institutes of Health under Award Number R01LM011834.

	
	
	%
	\bibliographystyle{IEEEtran}
	\bibliography{ICHI_bib}

\end{document}